\pdfoutput=1
\documentclass[twocolumn]{emulateapj}
\usepackage{hyperref}

\shorttitle{Star Formation Quenching in Nearby Disk Galaxies}
\shortauthors{Guo et al.}

\begin{document}

\title{SDSS-IV MaNGA: The Roles of AGNs and Dynamical Processes in Star Formation Quenching in nearby Disk Galaxies}

\email{kxguo@pku.edu.cn, yjpeng@pku.edu.cn}

\author{Kexin Guo}
\affil{Kavli Institute for Astronomy and Astrophysics, Peking University, 5 Yiheyuan Road, Haidian District, Beijing 100871, P.R.China}
\affil{International Centre for Radio Astronomy Research (ICRAR), University of Western Australia, Crawley, WA 6009, Australia}

\author{Yingjie Peng}
\affil{Kavli Institute for Astronomy and Astrophysics, Peking University, 5 Yiheyuan Road, Haidian District, Beijing 100871, P.R.China}

\author{Li Shao}
\affil{Kavli Institute for Astronomy and Astrophysics, Peking University, 5 Yiheyuan Road, Haidian District, Beijing 100871, P.R.China}

\author{Hai Fu}
\affil{Department of Physics and Astronomy, University of Iowa, Van Allen Hall, Iowa City, IA 52242, USA}

\author{Barbara Catinella}
\affil{International Centre for Radio Astronomy Research (ICRAR), University of Western Australia, Crawley, WA 6009, Australia}
\affil{ARC Centre of Excellence for All Sky Astrophysics in 3 Dimensions (ASTRO 3D)}

\author{Luca Cortese}
\affil{International Centre for Radio Astronomy Research (ICRAR), University of Western Australia, Crawley, WA 6009, Australia}
\affil{ARC Centre of Excellence for All Sky Astrophysics in 3 Dimensions (ASTRO 3D), Australia}

\author{Feng Yuan}
\affil{Key Laboratory for Research in Galaxies and Cosmology, Shanghai Astronomical Observatory, Chinese Academy of Sciences, 
80 Nandan Road, Shanghai 200030, China}

\author{Renbin Yan}
\affil{Department of Physics and Astronomy, University of Kentucky, 505 Rose Street, Lexington, KY 40506-0055, USA}

\author{Chengpeng Zhang}
\affil{Kavli Institute for Astronomy and Astrophysics, Peking University, 5 Yiheyuan Road, Haidian District, Beijing 100871, P.R.China}

\author{Jing Dou}
\affil{Kavli Institute for Astronomy and Astrophysics, Peking University, 5 Yiheyuan Road, Haidian District, Beijing 100871, P.R.China}

\begin{abstract}

We study how star formation (SF) is quenched in low-redshift disk galaxies with integral-field spectroscopy.
We select 131 face-on spiral galaxies with stellar mass greater than $\rm 3\times10^{10}M_\odot$, 
and with spatially resolved spectrum from MaNGA DR13.
We subdivide the sample into four groups based on the offset of their global specific star formation rate (SFR)
from the star-forming main sequence and stack the radial profiles of stellar mass and SFR.
By comparing the stacked profiles of quiescent and star-forming disk galaxies,
we find that the decrease of the global SFR is caused by the suppression of SF at all radii,
but with a more significant drop from the center to the outer regions following an inside-out pattern.
As the global specific SFR decreases, the central stellar mass, the fraction of disk galaxies hosting stellar bars, and 
active galactic nuclei (AGNs; including both LINERs and Seyferts) all increase, 
indicating dynamical processes and AGN feedback are possible contributors to the inside-out quenching of SF in the local universe.
However, if we include only Seyferts, or AGNs with ${\rm EW(H\alpha)>3\AA}$,
the increasing trend of AGN fraction with decreasing global sSFR disappears.
Therefore, if AGN feedback is contributing to quenching,
we suspect that it operates in the low-luminosity AGN mode,
as indicated by the increasing large bulge mass of the more passive disk galaxies.

\end{abstract}


\section{Introduction}

Galaxies in the local universe are clearly separated into two populations in the color-magnitude diagram (CMD), 
``Red Sequence'' and ``Blue Cloud,'' with only few galaxies scattering between forming a ``Green Valley (GV).''
This bimodality in distribution of galaxy optical properties in terms of their star formation (SF) activity, 
is found to be not only related with evolutionary stage, but also closely correlated with morphology of a galaxy \citep[e.g.,][]{K03b, Wuyts11}.
Star forming galaxies (SFGs) are dominated by spiral disks, 
while spheroidal galaxies show little or no emission generated from the H{\scriptsize\,II} region ionized by young stars.
Any mechanism that changes galaxies into `red and dead' (i.e., quenching) is required 
to explain the morphological transformation in a self-consistent way.

What causes the quenching of a galaxy has long been debated. 
Many factors are proven to be important in suppressing SF but 
responsible mechanisms are not exactly the same in different galaxies.
Following the reddening direction of CMD, the mode of historical SF in galaxies changes from a long-timescale pattern 
into a major merger triggered starburst \citep{S15},
suggesting an evolution in quenching mode from early to late cosmic epoch \citep{HC16}.
Theoretical and numerical works require active galactic nuclei (AGNs) feedback 
in stopping SF to form local massive early-type galaxies \citep[][and references therein]{E17}, 
which are quenched $\sim 10\,{\rm Gyr}$ ago \citep{Kauffmann03},
with major mergers invoked to explain their rapid star formation history (SFH) and spheroidal morphology \citep{S15}.
Mergers are also assumed in simulation to form classical bulge in spiral galaxies \citep[][and references therein]{BC16}.
However, given the low merger rate in the local universe \citep{Lotz11}, 
the build-up of spheroidal structure in the galaxy center could be attributed to other physical reasons rather than mergers.
Dynamical processes during secular evolution are considered to be relatively more important in 
driving (pseudo) bulge formation and passive evolution in present-day galaxies \citep[][and references therein]{Gadotti11, Kruk18}.

Mass-dependent long-timescale \citep[$\sim$4\,Gyr as suggested by][]{Peng15} evolution for SFGs is suggested 
by the finding of the star formation main sequence \citep[SFMS, e.g.][]{Speagle14},
with the star formation rate (SFR) of most SFGs strongly correlating with their stellar mass 
following a tight log-linear relation.
The flattening in slope of the local SFMS found in the massive end \citep[e.g.]{Lee15} suggests 
a mass-related global SF suppression for massive galaxies compared to lower-mass disks, 
with significant non-starforming bulge/central structure build-up in massive spirals argued to be 
the most important reason \citep{Abramson14,Bluck14,Whitaker17}.
Using the position of galaxies on the SFR-$M_*$ diagram with respect to SFMS as 
a tracer of the current evolutionary stage of galaxies,
it have also been found that normal SFGs are almost exponential disks, 
while starbursts and galaxies fading out of SFMS are observed to be highly concentrated \citep{Wuyts11,Guo15,Morselli17}.

``Morphological quenching'' is suggested as a morphologically related mechanism 
that stops SF in massive galaxies,
with the existence of prominent bulge increases the Toomre-{\it Q} parameter \citep{Toomre64}, 
and ``stabilizes'' gas on disks, i.e., prevents cold gas from collapsing and forming stars \citep{Martig09}.
While this ``morphological quenching'' naturally leads to an ``inside-out'' quenching pattern 
suggested by observation of local massive galaxies \citep{Pan15,B17b},
an alternative inside-out quenching mechanism is also pointed out by observational studies based on 
mass profiles of galaxies at different evolutionary stages \citep{Tacchella15,Tacchella17},
with a compaction followed by starburst and gas depletion in galaxy center suggested to be 
necessary in suppressing SF \citep{Barro15, Whitaker17}.
The latter mechanism always requires a dynamical process to trigger gas in the galactic center,
to fuel the upcoming intensive central SF and the possible accompanying stellar or AGN feedback.

On the other hand, the finding that galaxies in transition between star forming and quiescent 
are central low ionization emission-line regions \citep[][see also \citealt{Bar17} for AGN hosts above SFMS]{B17}
directly suggests the feedback of low-luminosity galactic nuclei (LLAGN) as another possible reason of galaxy quenching.
However, whether the AGN activity directly correlates with SF intensity or 
whether the AGN activity is related with the build-up of the central structure in disk galaxies is still not fully understood.

In this paper, we examine the resolved SF properties of massive spiral galaxies crossing the SFMS downward, 
to statistically explore how stellar mass is assembled during the passive evolution of a galaxy and how does AGN activity relates with 
the evolutionary stage of SF, 
aiming to find out if a plausible AGN activity is related with the galaxy structure build-up by comparing 
the spatial distribution of emission-line ratios with that of SFR and stellar mass ($M_*$).
Obviously there is no direct evolutionary link between current SFGs and quiescent galaxies, 
with the latter believed to have had active SF in an epoch as early as $z\sim0.5$ \citep{Peng15}.
However, we can still explore the crucial mechanisms that cause the quenching of a galaxy by 
comparing present-day galaxies in different evolutionary stages,
given that passive galaxies at fixed stellar mass had been sharing the same properties in both morphology and SF activity
with SFGs until they faded out of SFMS.

This study is based on the MaNGA (Mapping Nearby Galaxies at Apache Point Observatory, \citet{Bundy15}) public database 
from the 13th Data Release of the Sloan Digital Sky Survey \citep[SDSS-DR13;][]{DR13}. 
Only massive ($M_*>10^{10.5}\,\rm M_\odot$) spiral galaxies are examined for the reasons that:\\
(1) the evolution of massive galaxies is less affected by environmental effects;\\
(2) quenching and any morphological transformation during quenching events for these galaxies are unlikely to be caused by major mergers.

We describe the data analysis in the next section, and show the resolved properties of emission lines and the stellar component in Section \S3. 
We summarize and discuss our results in Section \S4.
A \citet{Chabrier03} initial mass function (IMF) is used throughout this work unless otherwise stated. 
We assume the following cosmological parameters: $\Omega_0=0.3,\Omega_\Lambda=0.7$, and $H_0=70\,\rm {km s^{-1}}$.

\section{Sample and Data analysis}
MaNGA is the largest ongoing integral-field unit (IFU) survey of low-redshift galaxies, 
designed to explore the internal kinematic structure and composition of gas and stars in 10,000 nearby galaxies
with a spectral coverage from $3600$ to $10300\rm\AA$ at a typical resolution $R\sim2000$ \citep{Bundy15}. 
Galaxies are covered to at least $1.5\, R_{\rm e}$ in observation.
MaNGA target galaxies are selected from the NASA-Sloan Atlas catalog \citep[NSA;][]{Blanton11}.
The selection has a wide coverage in stellar mass ($M_*>10^9\,M_\odot$) and optical color, 
to allow an unbiased analysis of SF of the galaxy population at different evolutionary stages.
The median point-spread function (PSF) of the MaNGA data cubes has a median full width at half maximum (FWHM) of $2\farcs5$ \citep{Law16},
with the standard deviation of $\sim0\farcs1$ in $r$-band.

We utilize the visually classified morphology from GalaxyZoo \citep{Lintott08,Lintott11}.
From 1272 matches found in 1390 galaxies in SDSS-DR13/MaNGA database with a matching radius of 1'', 
131 spirals with $M_*>10^{10.5}\rm M_\odot$ are selected.
The stellar masses are taken from the NSA\,${\rm v1\_5\_4}$ catalog 
\footnote{We divide the stellar mass in the NSA catalog by a factor of 0.49 to keep the consistency of cosmology.}.
We have avoided highly inclined galaxies by removing those with axis ratio $b/a<0.5$.
These galaxies are distributed in the redshift range $0.02<z<0.14$, 
with a median value $z_{\rm med}=0.05$.
118 (90\%) galaxies in our sample are distributed within $0.02<z<0.08$.

\begin{figure*}
\plotone{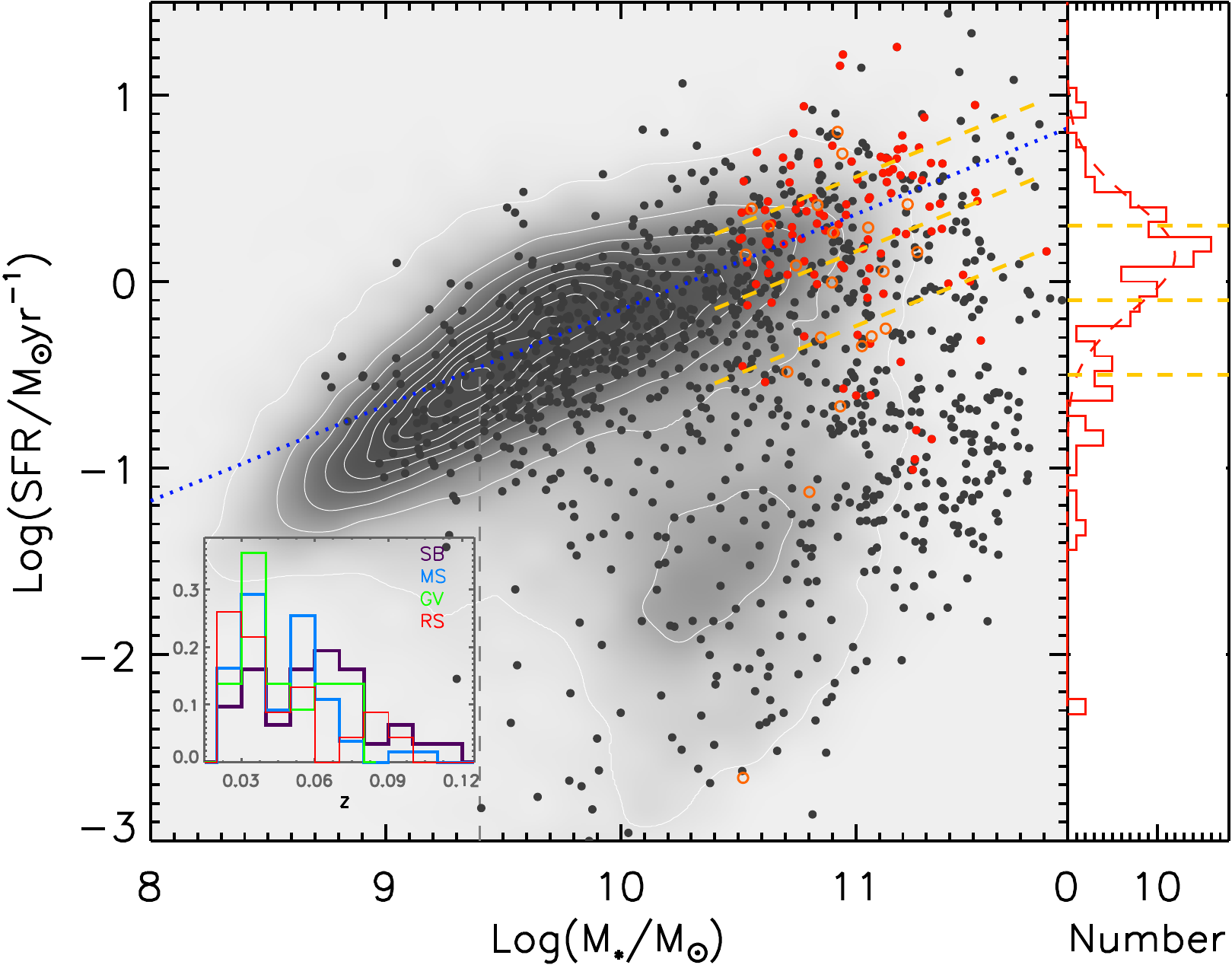}
\caption{Relation between integrated SFR and $M_*$ of SDSS-DR13/MaNGA galaxies (black dots, 
only galaxies matching GSWLC-X1 are shown). Red dots and orange circles indicate our target face-on spirals 
with and without GSWLC counterparts, respectively. 
For the latter, integrated SFR from $\rm H\alpha$ (see \S3.1) is used.
The distribution of GSWLC galaxies with the same redshift and stellar mass distribution of SDSS-DR13/MaNGA galaxies is shown by the greyscale background and contours,
with the vertical gray dashed line indicating a completeness limit in stellar mass.
The right panel shows the SFR distribution of the massive face-on spirals (red+orange) relative to SFMS (blue dotted line), 
normalized to the corresponding SFR at $\rm log(M_*/M_\odot)=10.5$.
Yellow dashed lines in either panel show the separations between Starbursts ({\bf SB}s), Main Sequence galaxies ({\bf MS}s), 
Green Valley galaxies ({\bf GV}s) and Red Spiral galaxies ({\bf RS}s, see \S3.1).
The lower-left inner panel shows the normalized redshift distribution of galaxies in the four classes.
\label{fig1}}
\end{figure*}

The data cubes of our targets are rebinned to $1''\times 1''$ pixels, 
to remove the covariances between neighboring pixels. 
The rebinning also speeds up the spectral fitting afterward.
A map of the signal-to-noise ratio (S/N) calculated in the $\rm H\alpha$ region (rest-frame $6525-6610\rm\AA$) is made for each galaxy.
Voronoi binning \citep{CC03} is then applied to those pixels with S/Ns less than 30.
The covariance in pixel binning is accounted for simply by applying
\begin{equation}
\rm noise_{covar}/noise_{no\_covar}=1+1.62\times\log(N_{bin}), 
\end{equation}
following the online instructions of MaNGA data reduction\footnote{http://www.sdss.org/dr13/manga/manga-caveats/}. 
The foreground galactic extinction is estimated based on the dust map of \citet{SFD98},
and is corrected using the Galactic extinction law of \citet{CCM} with $R_V=3.1$. 

The spectral fitting pipeline LZIFU \citep{Ho16} is utilized to analyze stellar component and emission lines.
Stellar continuum is fit and extracted using the ``penalized pixel-fitting'' (pPXF) routine \citep{ppxf04,ppxf17},
before fitting Gaussian line profiles to emission lines.
Weighted simple-stellar-population (SSP) templates from MIUSCAT \citep{Vazdekis12} are applied in continuum fitting for the entire wavelength coverage.
A Salpeter IMF is used
\footnote{SSP spectral energy distributions (SEDs) of metallicity $\rm [Z/H]=-0.71, -0.40, 0, +0.22$ 
and age from 63\,Myr to 18\,Gyr (0.2 dex steps)
with unimodal IMF of slope 1.3 (i.e., Salpeter case) are used in continuum fitting. 
All resolved properties are based on Salpeter IMF to keep the consistency.
Later in the text, 
the integrated $\rm H\alpha$-based SFR is corrected to Chabrier IMF to compare with that from the GSWLC-X1 catalog.}.
The stellar mass for each pixel is estimated by summing up the weights of the best-fit stellar templates.
Pixels that belong to other sources in the field of view rather than target galaxies 
have been selected and masked based on segmentation maps reduced from SDSS/$r$-band images.
We then classify the pixels in terms of their position on the \citet[][BPT]{BPT} diagram, 
by comparing their [O{\scriptsize\,III}]/$\rm H\beta$ and [N{\scriptsize\,II}]/$\rm H\alpha$ 
with those of SFGs and AGNs, following \citet{K06} and adopting the empirical separation between Seyfert and LINER defined by \citet{S07}. 
Thus pixels of each galaxy are classified into {\it Star-forming, Composite, LINER-like} and {\it Seyfert-like} regions.
We refer to \citet{B04} for the classification of pixels with $\rm 1<S/N<3$ in any lines.
Particularly, 
we assign pixels which have [N{\scriptsize\,II}]/$\rm H\alpha>0.6$ and $\rm S/N\ge 3$ in both lines 
but $\rm S/N<3$ in $\rm H\beta$ or [O{\scriptsize\,III}]
to be LINER-like regions. 

\section{Methods and Results}
The position of each galaxy on the integrated $\rm SFR$- $M_*$ plot (Figure~\ref{fig1})
indicates their global evolutionary stage.
Galaxies are classified into Starbursts, normal SFGs, GV galaxies and Red Sequence galaxies
according to their integrated $\rm SFR$ and their distance to the SFMS at the same stellar mass.
To explore the mechanisms that suppress the the global SFR,
the resolved properties of the galaxies including stellar mass, SFR, light-weighted stellar age 
(indicated by Dn4000) and dominated ionization mechanism are analyzed.

\begin{figure}
\plotone{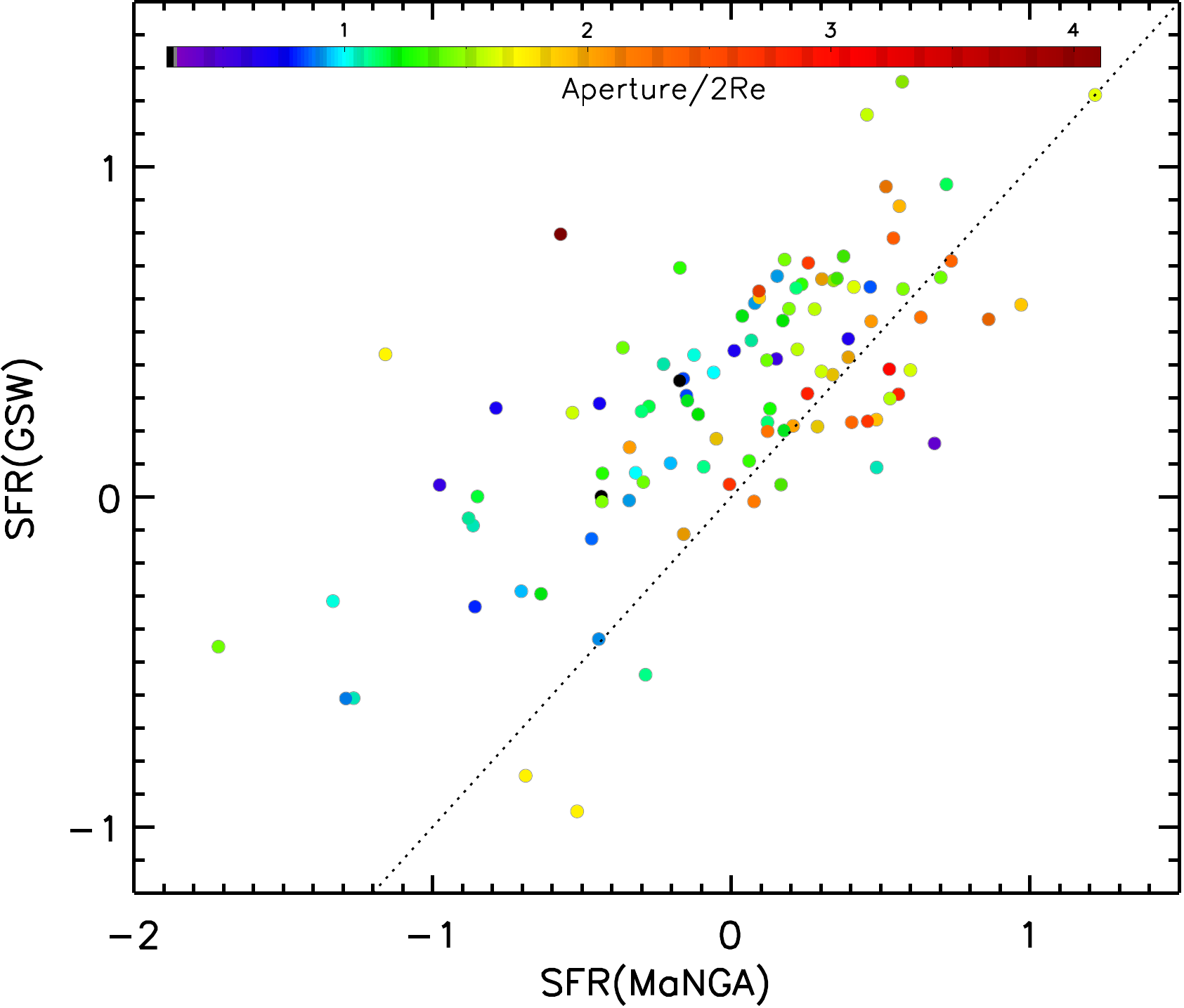}
\caption{The comparison of SFR from $GALEX$-SDSS-WISE Legacy Catalog (GSWLC) with 
that from integrated ${\rm H\alpha}$ emission in all available pixels within an aperture,
for the 115 matched galaxies.
Different color labels the relative size of an IFU aperture to the half-light radius ({\tt ``nsa\_sersic\_th50''}) of a galaxy.
The dotted line shows a one-to-one relationship.
MaNGA-based integrated SFR is on average smaller than that from GSWLC.
In general the larger the aperture is, the closer the two measurements are to each other.
In this paper we use the GSWLC one to classify the global evolutionary stage of galaxies.
\label{fig2}}
\end{figure}

\subsection{SF phase classification}
Given the relatively high sample completeness and accuracy in SFR estimation,
we use $GALEX$-SDSS-WISE Legacy Catalog \citep[GSWLC-X1,][]{Salim16} as the reference catalog to construct SFMS in the redshift range of our MaNGA galaxies.
The SFR estimation in this catalog is based on broadband photometry. 
It has an advantage of being less affected by AGN-heated dust emission,
and does not suffer from uncertainties of aperture correction.
The SFMS constructed based on the data above
\footnote{To construct the SFMS, 
we select galaxies in the GSWLC-X1 catalog with the same redshift distribution of galaxies in NSA ${\rm v1\_5\_4}$ catalog.} 
is given by:
\begin{equation}\label{eq:sfms}
 \log(\rm SFR/(M_\odot yr^{-1}))=0.51\times\log(\rm M_*/M_\odot)-5.28
\end{equation}
indicated by the blue dotted line in Figure~\ref{fig1}. 
This is determined by a log-linear fit for galaxies with $M_*>10^{9.4}\rm \,M_\odot$ 
which construct a gaussian-like SFR distribution above the separation between SFGs and passive sequence.

We define galaxies with SFRs $0.2\,\rm dex$ above the SFMS as ``Starburst'' ({\bf SB}) galaxies;
galaxies with SFRs between SFMS$+0.2\,\rm dex$ and SFMS$-0.2\,\rm dex$ as Main Sequence ({\bf MS}) galaxies;
galaxies with SFRs between SFMS$-0.2\,\rm dex$ and SFMS$-0.6\,\rm dex$ as {\bf GV} galaxies;
and those with SFRs below SFMS$-0.6\,\rm dex$ as passive spiral galaxies ({\bf RSs}).
It should be noted that a more conventional definition of {\bf SB} galaxies ({\bf RS} galaxies) would require a higher (lower) cut of SFR relative to the SFMS.
This will, however, give too few galaxies classified as SB galaxies and RS galaxies, due to the relatively small size of our sample.
Therefore, we adopt the above operational definition of each class.

We cross-match the NSA\,${\rm v1\_5\_4}$ catalog with GSWLC-X1 using $1''$ as matching radius,
\footnote{12 more matches would be found if a $3''$ matching radius was used.
However, these galaxies are either interacting/diffuse or highly inclined, 
hence we did not include these galaxies in our final sample.}
and adopt SFRs from GSWLC catalog for the matched sample.
Within the selected 131 massive spiral galaxies in our MaNGA sample, 115 of them have matches.
For the other 16 galaxies that do not have matches in the GSWLC-X1 catalog,
we calculate their SFR by summing up all the dust-corrected $\rm H\alpha$ emission from star-forming regions 
defined in \S2 using the formula from \citet{K98}:
\begin{equation}
{\rm SFR}=7.9\times10^{-42} L_{\rm {H\alpha,corr}},
\end{equation}
which should be considered as a lower limit for the reason of both ignoring SF from other gas-ionized regions 
and the limited radial coverage of MaNGA observation.
Dust extinction is derived from the observed Blamer decrement with intrinsic $\rm H\alpha/H\beta=2.86$, 
using \citet{CCM} extinction law applied with $R_V=3.1$ and other coefficients updated by \citet{OD}.
We have compared the SFR from the GSWLC catalog and that from integrated ${\rm H\alpha}$ emission for the matched galaxies,
which is shown in Figure~\ref{fig2}. 
Because the nontrivial aperture correction is out of the scope of this article and, 
MaNGA-based SFR is on average smaller than that from GSWLC,
we decide to use the latter in classifying the global evolutionary stage of a galaxy.
The global $\rm H\alpha$-based SFR is divided by a factor of 1.7 to be compared to the Chabrier IMF-derived SFR in GSWLC.

By comparing the position of the 131 face-on massive spirals of our MaNGA galaxy sample with SFMS defined in equation\,(\ref{eq:sfms})\,(see Figure\,\ref{fig1}), 
31 galaxies are classified as {\bf SB}s, 55 are {\bf MS}s,
22 are in {\bf GV}s and 23 are {\bf RS}s. 
Though the SFR measurement based on emission lines could be inconsistent with that derived from broadband photometry, 
the removal of the 16 galaxies with only MaNGA detection does not affect the 
average distribution of SF or stellar mass for galaxies in different evolutionary stages.
We stress again that {\bf SB}s defined here are those galaxies with integrated SFR 0.2\,dex higher than those on SFMS.
They do not show any merging features following our visual identification, 
and could be classified into ``Main Sequence'' if a more conventional definition of SBs with SFR 4 times higher than the MS galaxies.
{\bf RS} is also not necessarily to the same ``Red Sequence'' as that commonly defined by the literature,
which is dominated by fully quenched early-type galaxies about 1\,dex below the {\bf MS}.
We remind our readers that all these galaxies are ``disks''
(including those showing clear disks but no distinct spiral arms) in morphology.

\subsection{Stacking and AGN contamination}
To give a clear and global view of the resolved properties of galaxy population 
in the four different SF phases as defined above,
we construct the maps of median $M_*$, SFR, sSFR, Dn4000 and BPT classification 
by stacking the galaxies in each SF phase and taking the median values of each stacked pixel, i.e., median stacking.
We also construct the one-dimensional (radial) median stacking of SFR, sSFR and $M_*$ which gives a more quantitative way of exploring 
the evolution in $\sim$kpc/$\sim0.5\,R_{\rm e}$ scale.

\begin{figure*}
\plotone{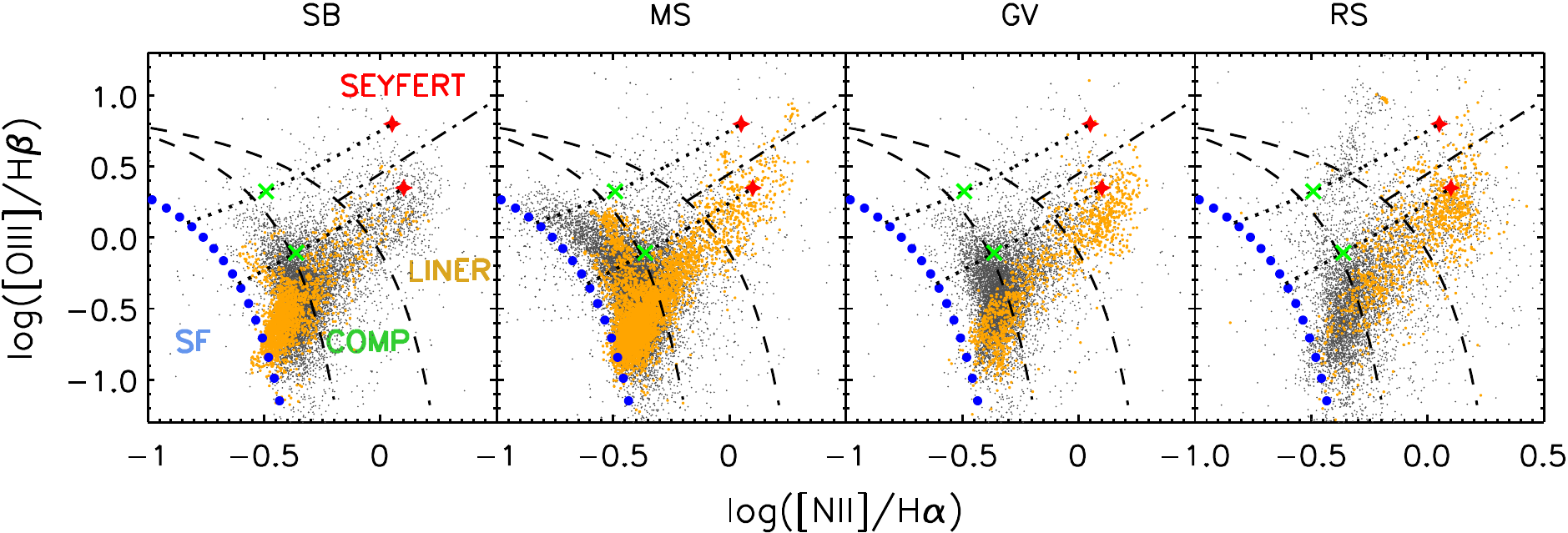}
\caption{The classification of all pixels (black dots) in terms of their position on BPT diagram 
($\rm{log([N}${\scriptsize\,II}$\rm{]/H\alpha)}$ vs. $\rm{log([O}${\scriptsize\,III}$\rm{]/H\beta)}$), 
separation between different types utilized by \citet{K06} and \citet{S07} are applied.
Pixels within 1\,$R_{\rm e}$ are colored in orange.
star-forming, composite, Liner-like and Seyfert-like regions are separated by dashed lines.
The two red star symbols in each panel are 
templates of star-forming galaxies and AGNs adapted from \citet{KK09},
while the blue dotted line is a shift of \citet{K03} SF/AGN border to make sure the 
separation between SF and composite to be with 50\% SF/AGN contribution.
The black dotted lines show two random examples for AGN-SF decomposition described in \S3.2, 
where X symbols indicate the position on the BPT diagram when either side contributes $50\%$ to observed $\rm H\alpha$ flux.
We notice that galaxies, especially the central regions, are dominated by star formation in ``Starburst'' and ``Main Sequence'', 
becoming more composite and LINER-like when they transit to the other two populations.
\label{fig3}}
\end{figure*}

We ignored the $\rm H\alpha$ emission from pixels that are not classified as SF regions 
in calculating global SFR of individual galaxies.
However, simply masking out higher-ionized regions leads to an inaccuracy in stacked SFR.
This could be either underestimated because of not accounting SF in LINER- or Seyfert like regions,
or overestimated by ignoring the un-SF pixels in stacking,
due to the low $\rm H\alpha$ contribution from LINER-like pixels compared to those dominated by H{\scriptsize\,II} regions,
unless SFR of 0 is assigned to AGN-like pixels and accounted for. 
Thus we turn to the AGN-SF decomposition method in \citet{KK09} to refine the SFR from individual pixels 
according to their position on the BPT diagram.
This method offers a statistically reliable estimation of averaged SFR, 
and the contribution from either side does not suffer from the uncertainty caused by 
the difference in intrinsic Balmer decrement 
between AGN-like and H{\scriptsize\,II} regions when extinction correction is applied.
We apply this method to pixels that are classified as 
composite, LINER-like, and Seyfert-like in Section\,\S2.
The two ``pure-AGN'' templates based on SDSS spectra used in \citet{KK09} are adopted.
In addition,
we adopt a shift of the separation line between SFG and AGN from \citet{K03} to represent the ``Star forming'' ridge,
so that the SF/composite border is around 50\% SF contribution.
The classification of ``Star forming'' and ``AGN-like'' regions together with the templates used in decomposition are shown in Figure 3.

An S/N cut is applied to pixels before stacking. 
Since we aim to explore the galaxy properties as a function of distance to galactic center,
given that the mass and light distribution of a galaxy generally follows a radially symmetric pattern,
we apply a contour binning \citep{contbin} to pixels with S/N in $\rm H\beta\le 1$ for each galaxy before extinction correction.
A simple deprojection method is also applied to 2D maps of each bin before they are rescaled to physical scales and stacked.
In practice, we rotate a galaxy to align its major axis horizontally and 
elongate the image in the perpendicular direction to match the minor and the major axes,
and then we transform the image back to its original size.
Since our galaxies are selected to have inclination angles less than $60^{\circ}$ ($b/a<0.5$), 
the uncertainty caused by the deprojection process does not affect the stacking results.
Similarly, with the assumption that disks are circular,
radial distance from pixels to galaxy center for individual galaxies are corrected geometrically using their global axis ratio and position angle
provided in the NSA catalog.

\begin{figure*}
\plotone{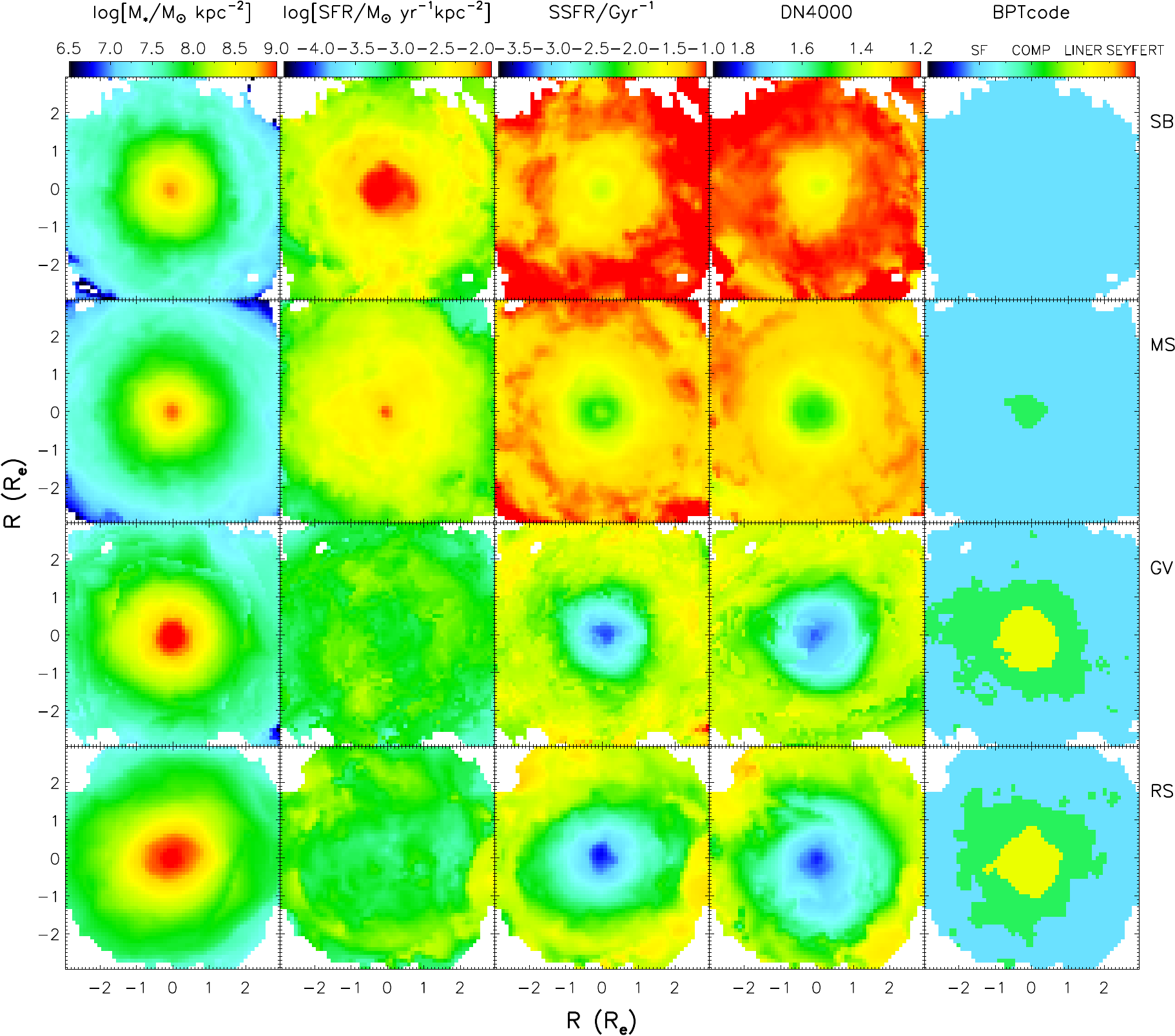}
\caption{Median stacked maps of $M*$,$\rm SFR$,$\rm SSFR$,Dn4000, and the position on BPT diagram of each evolution population.
Rows from the top down are the stacked maps for {\bf SB, MS, GV} and {\bf RS} respectively.
The ``BPT code'' labeled for the rightmost column is consistent with the color coding of Figure 3, 
corresponding to populations from star forming region to AGN-like pixels indicated by 
blue (H{\scriptsize\,II} region), green (composite), yellow (LINER-like), and red (Seyfert-like) areas.
The total stellar mass of individual galaxies has been normalized to $10^{10.5}\rm{M_\odot}$ to compare the distribution of surface density.
SFR here has also been set to the ``distance'' relative to SFMS as well.
It is obvious from this plot that, following the passive sequence indicated by the aging of central stellar population from the top down,
star formation in galaxies fades out, 
the pattern of which transits from a central-peak disk, into a ring-like structure with a dip in galaxy center.
It is noticeable that the central part of a galaxy appears ``LINER-like'' after it falls into {\bf GV} and {\bf RS}.
\label{fig4}}
\end{figure*}

\subsection{Resolved properties}
Figure~\ref{fig4} shows the median stacked maps of $M_*$, $\rm SFR$, $\rm SSFR$, Dn4000, 
and the emission-line classification based on the resolved BPT diagram in the panels from the left to the right, 
for the four populations from {\bf SB} to {\bf RS} indicated by panels from top to bottom.
Stacking is only applied to places where the number of available pixels are greater than 3 to make the median estimates reliable.
The typical FWHM of PSF is $2\farcs5$,
which corresponds to less than 3\,kpc, i.e., less than $1\,R_{\rm e}$ of most galaxies in our sample
\footnote{Most (90\%) of our galaxies are within the redshift range of $0.02<z<0.08$,
thus the smearing effect is limited in a scale less than $\sim3.2\,\rm kpc$, 
which is less than $1\,R_{\rm e}$ for $66\%$ of our sample.}.
Also the redshift distribution of the 4 galaxy groups is identical, as shown in Figure~\ref{fig1} (lower-left panel),
the relative difference of resolved properties between different groups without PSF deconvolution should not be affected by the smearing effect.
To test this, we have restricted the sample to galaxies with $10^{10.8}<M_*/\rm{M_\odot}<10^{11.2}$ in $0.02<z<0.08$.
The stacked results changed very little.

Figure 4 clearly demonstrates several key features of the resolved properties of the galaxy population,
and their changes as the integrated SFR decreases from {\bf SB} (top row) to {\bf RS} (bottom row).

\begin{enumerate}
\item As the integrated SFR drops from {\bf SB} to {\bf RS},
a clear increase in the central surface stellar mass density is seen.
This shows the clear bulge growth during the quenching process.

\item The suppression of SFR in {\bf GV} and {\bf RS} compared to the other two groups
is more significant in galaxy center than in the outer disk,
with the central-peak in $\rm\Sigma(SFR)$ distribution disappearing gradually with the passive evolution,
resulting in a ``ring-like'' SF indicated by our stacked {\bf GV} and {\bf RS} disks.

\item The aging of stellar population indicated by Dn4000 and 
the expanding of which from galaxy center to outer disk are very obvious.
This trend goes together with the globally fading of specific star formation rate (sSFR),
which shows a significant central-dip pattern since galaxies leave SFMS.

\item The rightmost column shows that disk center appears ``LINER-like'' in {\bf GV},
and stays like a low-luminosity AGN in {\bf RS} before the galaxy is fully quenched;
\end{enumerate}

We remind our readers that these are all median features. 
For example, the composite classification for the stacked central region of MS galaxies
does not necessarily mean that most of the MS galaxies have a transition-type AGN.
Actually only 7 of them have composite centers, while 27/21 in 55 are classified as star-forming/AGN-like.

\begin{figure*}
\minipage{0.32\textwidth}
  \includegraphics[width=\linewidth]{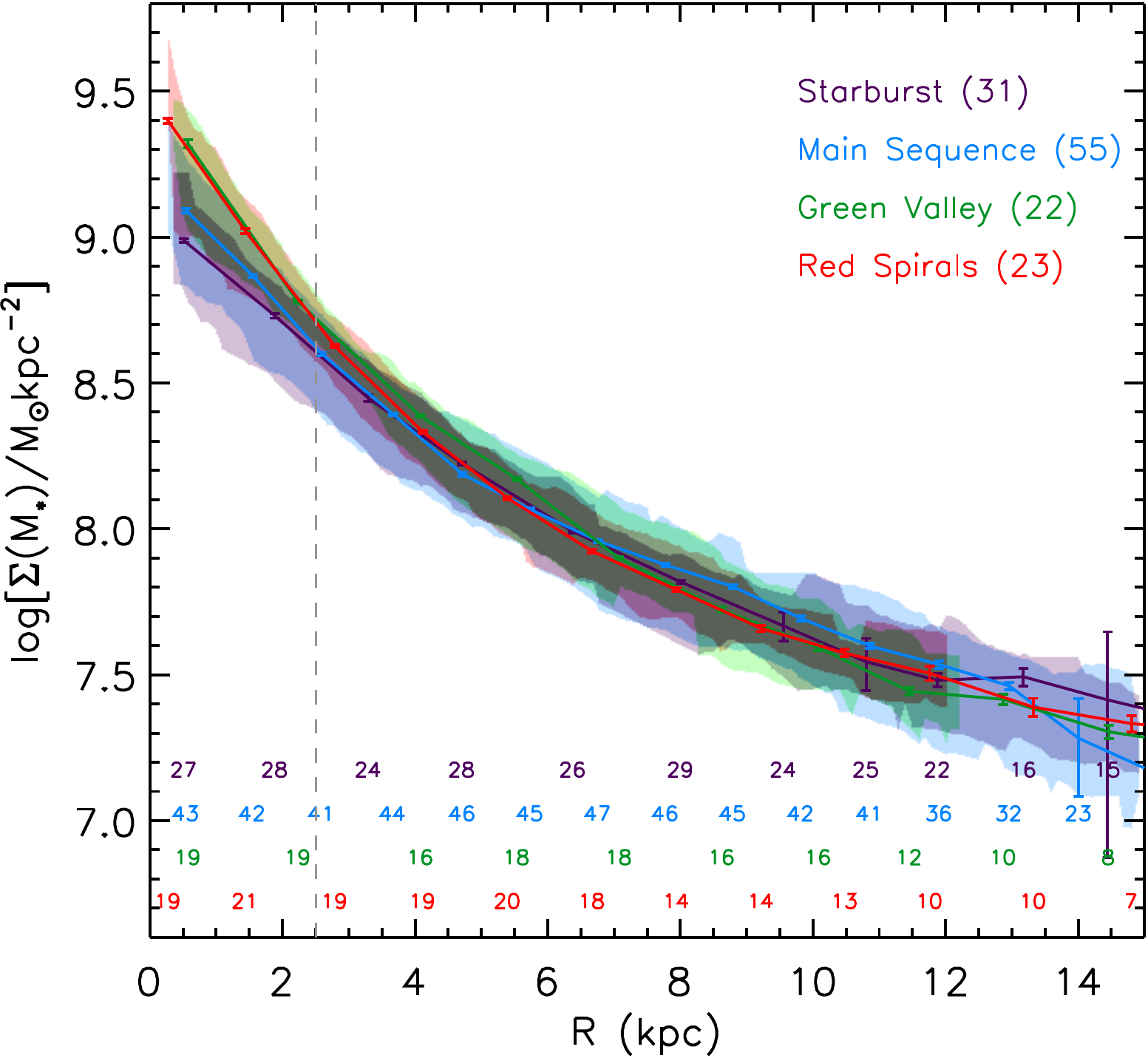}
\endminipage\hfill
\minipage{0.32\textwidth}
  \includegraphics[width=\linewidth]{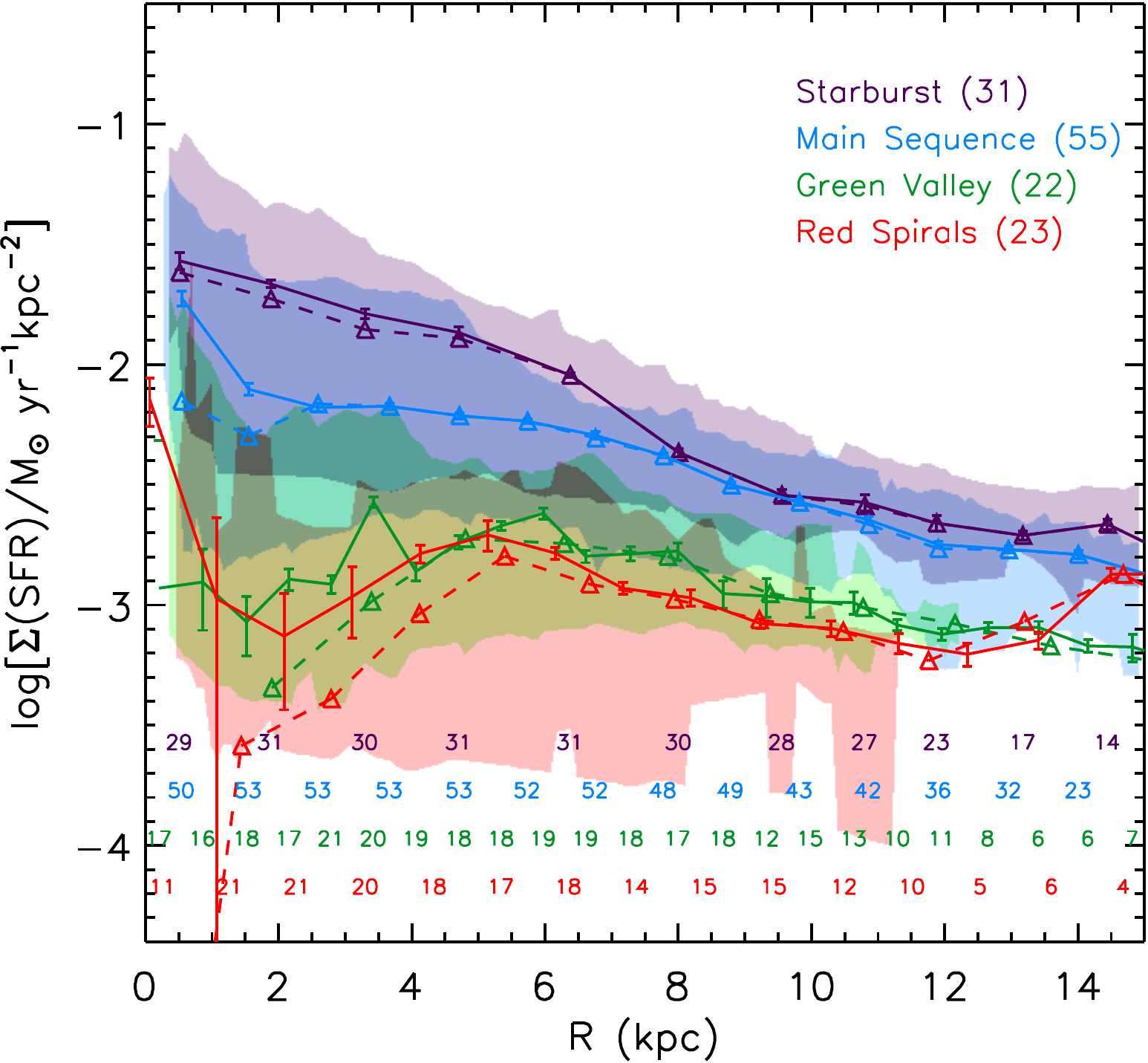}
\endminipage\hfill
\minipage{0.32\textwidth}
  \includegraphics[width=\linewidth]{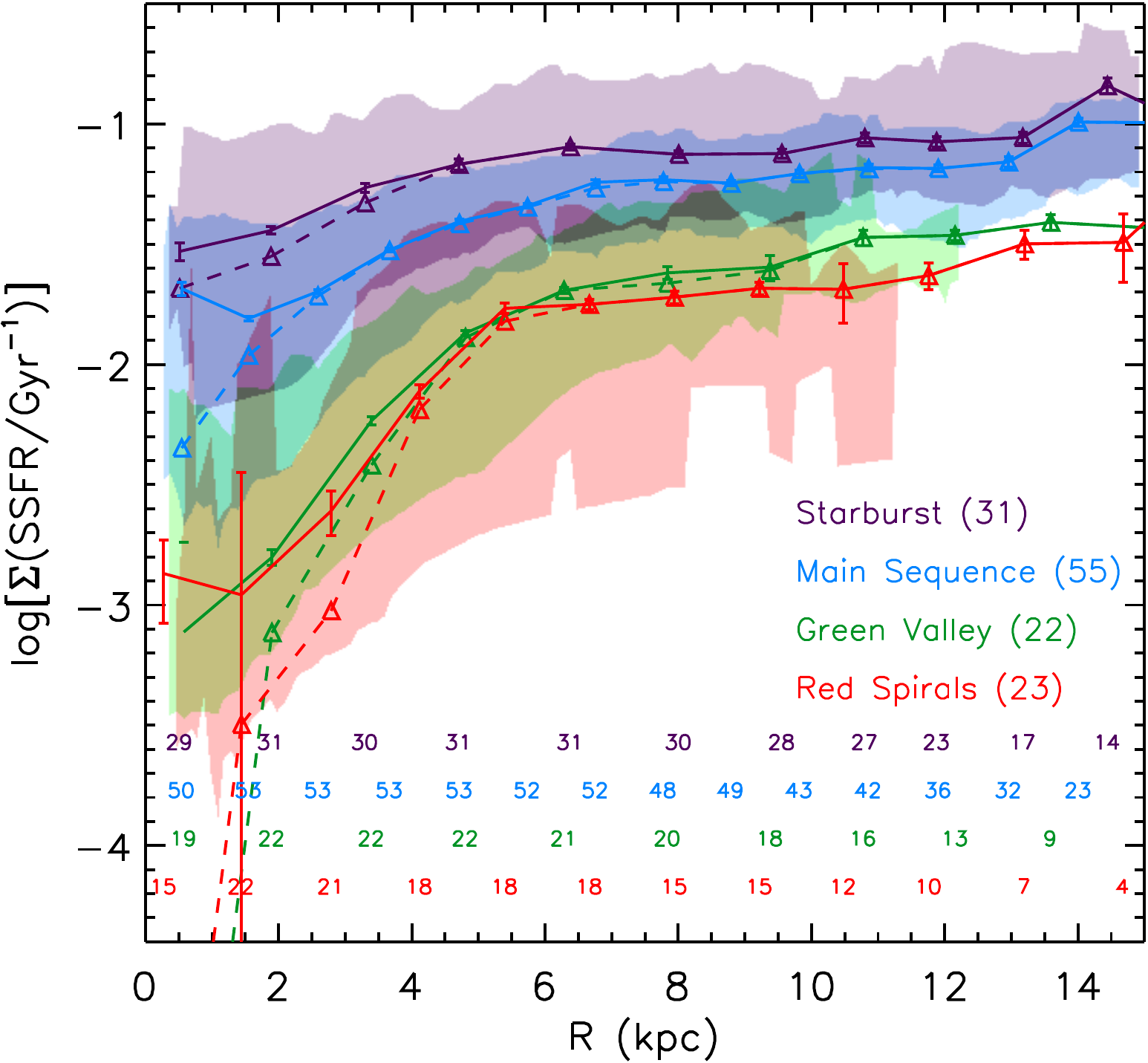}
\endminipage\hfill
\minipage{0.32\textwidth}
  \includegraphics[width=\linewidth]{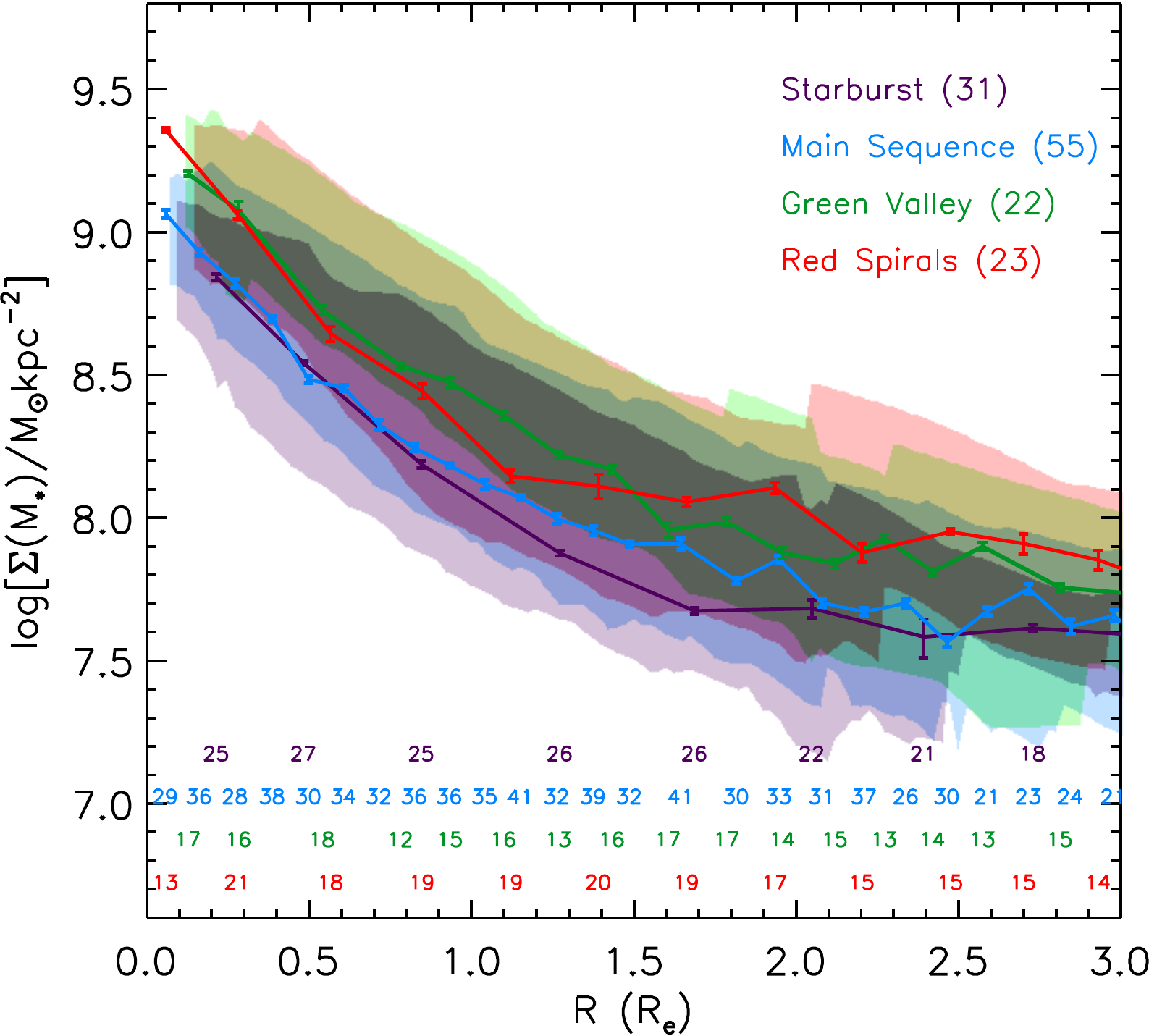}
\endminipage\hfill
\minipage{0.32\textwidth}
  \includegraphics[width=\linewidth]{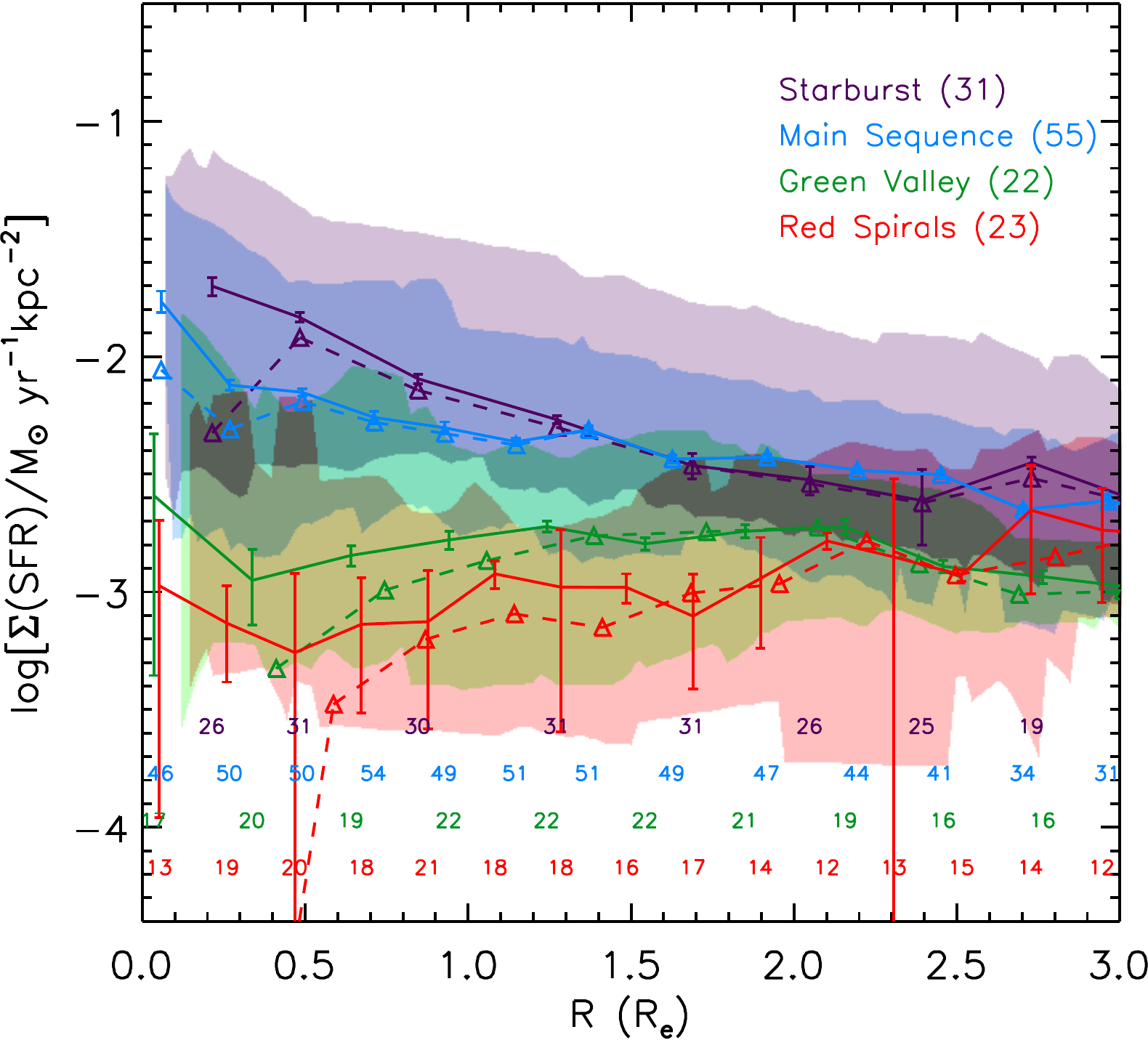}
\endminipage\hfill
\minipage{0.32\textwidth}
  \includegraphics[width=\linewidth]{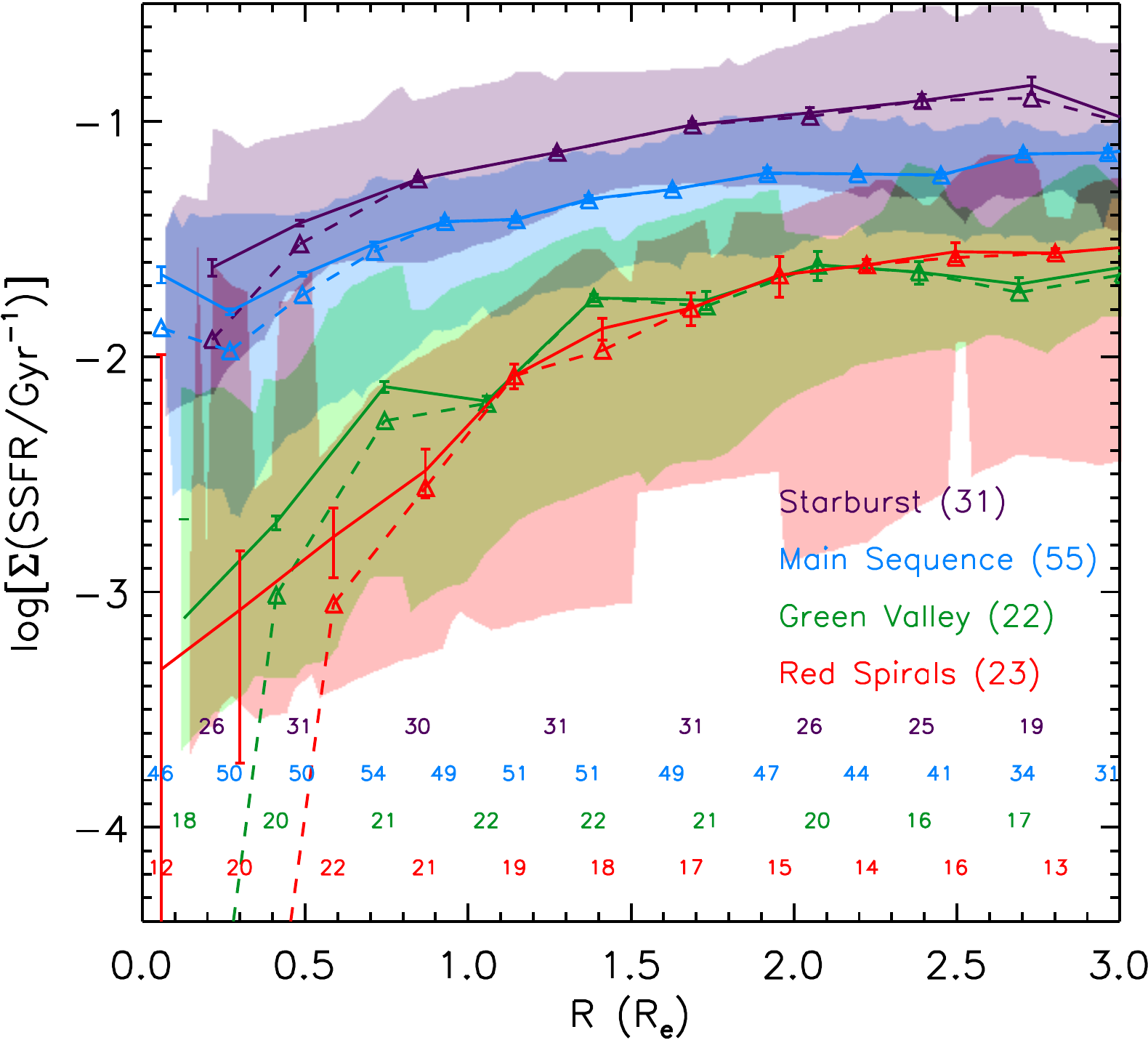}
\endminipage\hfill
\caption{From Left to Right panels: Stacked 1D surface density profiles for stellar mass, SFR, and SSFR respectively.
Numbers of stacked galaxies in each group are labeled in parentheses. 
The error bar for each data point is the average combined error of stacked values in each bin.
The same normalization as Figure~\ref{fig4} is applied to $\Sigma({M_*})$ and $\Sigma(\rm{SFR})$ before stacking.
The top and bottom panels show profiles in units of kiloparsec and $R_{\rm e}$ respectively.
Dashed lines are profiles if we set SFR in composite or AGN-like pixels as 0 before stacking.
Shaded areas in each panel indicate the $1\sigma$ scatter of the profiles for different groups, 
limited to radial bins where the numbers of available galaxies (not data points) are greater than 10.
The numbers of stacked galaxies in each bin are labeled above the $x$ axis.
We estimate the increase of central stellar mass within a 2.5\,kpc radius (vertical dashed gray line).
The decreasing of SFR is most noticeable in galaxy center compared to the global fading of disks,
during the build-up of central stellar mass from {\bf MS} to {\bf RS}.
\label{fig5}}
\end{figure*}

To give a quantitative comparison on the subgalactic scale, 
we apply a 1D median stack of the $\Sigma{(M_*)}$ and the $\rm{\Sigma(SFR)}$ profiles along the radial direction
with the IDL procedure {\tt regroup.pro}.
The $1\sigma$ scatter around the median value is shown as the shaded area in Figure~\ref{fig5}.
Both profiles are normalized to make better comparisons.
${\Sigma(M_*)}$ profiles are all normalized to an integrated $M_*$ of $10^{10.5} M_\odot$,
while $\rm{\Sigma(SFR)}$ profiles are normalized to the value of SFR on SFMS for corresponding $M_*$.
$\rm{\Sigma(SSFR)}$ profiles are derived from the normalized $M_*$ and $\rm SFR$ profiles.
A simple approach has been applied to test the beam smearing effect on the stacked profiles, as follows.

The PSF is assumed to be a Gaussian with an FWHM of $2\farcs5$.
Individual ${\Sigma(M_*)}$ profiles are fitted by a sum of four central peaked half-Gaussian functions 
with width and amplitude as free parameters
\footnote{We found that in practice, 
4 Guass is the simplest and best fit when multi Gaussian distributions are used in profile fitting.}.
The fitting is required to produce the same total integrated $M_*$ 
as the total $M_*$ derived from broadband photometry in the NSA catalog.
We then subtract the $2\farcs5$-FWHM gaussian profile from the stacked profile and find little difference.
We have not done this test to the $\rm\Sigma(SFR)$,
due to the large uncertainty in SFR estimation in the galaxy central regions,
especially for {\bf GV} and {\bf RS} galaxies with possible AGN contamination, 
as shown in the right pannels of Figure~\ref{fig5}.
However, 
we do not expect significant change in stacked profiles after PSF deconvolution,
since the smearing effect is limited in a scale $\le1\, R_{\rm e}$ for most of our galaxies.

As shown in Figure~\ref{fig5},
both $\Sigma(M_*)$ and $\rm\Sigma(SFR)$ change significantly from {\bf SB} to {\bf RS} at all radii,
with the largest change in the central regions($\le $2\,kpc).
The central $M_*$ increases $\sim$0.4\,dex from {\bf SB} to {\bf RS} and 
most of the increase happens during the transition from {\bf MS} to {\bf GV} phase.
The percentage of stellar mass within the central 5\,kpc diameter region 
(vertical dashed line of the top-left panel of Figure~\ref{fig5}) 
of a galaxy increase from 
31\% and 37\% in SB and MS, to 51\% and 50\% in GV and RS, 
with standard deviations of 19\%, 18\%, 19\% and 28\% in the four groups, respectively.
This growth of the central stellar mass as the decrease of the global SFR 
is related to the central peaked SF of {\bf SB} and {\bf MS} galaxies.
While the suppression of the SF happens across the entire disk of the galaxy,
from the center to the outer regions,
the most significant decrease appears at galaxy center, 
especially if we set the SFR contribution from ``LINER-like'' pixels to 0 (indicated by dashed lines).
Toward the end of this ``inside-out'' quenching process,
the SF in the central regions has dropped about 1.5\,dex,
in terms of both $\rm\Sigma(SFR)$ and $\rm\Sigma(SSFR)$,
leaving some residual star formation in the outer disk regions and forming a ring-like structure,
which is clearly shown in 2D SFR and sSFR distribution for {\bf GV} and {\bf RS} in Figure~\ref{fig4}.

Our results can be understood as the following key evolutionary stages for galaxy population:\\
\begin{enumerate}
\item[(1)] {\bf SB}s have the youngest stellar population and the most flat stellar disks. 
Although the SFR profile is peaked in the central regions,
the sSFR profile is nearly flat, 
which indicates the SF is similarly active across the entire disk.
Galaxy center shows the strongest emission from the H{\scriptsize\,II} region compared to outer parts.
\item[(2)] For galaxies on the {\bf MS},
the SF activity becomes lower compared to {\bf SB} galaxies,
with a sign of more suppressed SF in the central regions.
Accordingly, the dominant stellar population in the galaxy center is getting older as revealed by the Dn4000 map in Figure~\ref{fig4}.
The emission at galaxy center is still dominated by the H{\scriptsize\,II} region.
\item[(3)] The dramatic decrease of SF at galaxy center changes the $\rm\Sigma(SFR)$ profile of a {\bf GV} galaxy into a central-dip function, 
with a ``ring-like'' structure appearing at $\sim$5\,kpc from galaxy center.
Stellar mass, on the other hand, has a prominent build-up in the central area relative to outer disks, 
which is related to the central-peak SF pattern in {\bf MS} galaxies.
The primary emission source at galaxy center is more ``LINER-like'' rather than SF.
\item[(4)] {\bf RS} galaxies show similar mass profiles as those of {\bf GV}s.
Both SFR and sSFR profiles continue to decrease from {\bf GV} to {\bf RS}, 
with a slightly larger decrease in the central regions.
The difference almost disappears at $\sim 2 R_{\rm e}$,
which causes the SFR and sSFR ring-like structure to become more visible in the 2D map as in Figure~\ref{fig4}.
These galaxies still show ``LINER-like'' emission, but only limited in 2\,kpc from galaxy center in most cases.
\end{enumerate}

Our results are quantitatively consistent with the recent study of \citet{Medling18} 
based on SAMI (Sydney-AAO Multi-object Integral-field unit) observations for spiral galaxies 
in the same $\rm \Delta(SFMS)$ and $M_*$ range.
Our stacked $\rm \Sigma(SFR)$ profiles for galaxies from {\bf SB} to {\bf GV} are also consistent with those in \citet{Ellison2018},
while the difference between the profiles of red galaxies could have resulted from our different treatments to AGN-like pixels at galaxy center.
Again, we need to caution that there is no direct evolutionary link between the four classes from {\bf SB} to {\bf RS},
i.e., the {\bf SB} or the {\bf MS} galaxies are not the progenitors of the {\bf RS} galaxies at the same redshift.
A full quenching process is not expected to finish during the redshift range covered by our sample,
given a typical quenching timescale of several gigayears.
Indeed, the change of the the $\Sigma(M_*)$ profile from {\bf MS} to {\bf GV} cannot be fully explained by
integrating the SFR profile over some reasonable quenching timescale.
In other words, the change of the $\Sigma(M_*)$ profile from {\bf MS} to {\bf GV} would require a long quenching timescale
($\rm >5\,Gyr$), or additional mass contribution from mergers or other dynamical processes that can help to increase the $\Sigma(M_*)$.
A more comprehensive analysis with consideration of quenching timescale,
progenitors at higher redshift, and mergers is necessary and will be discussed in our future work.

\section{Discussion and Conclusion}
Based on 131 face-on spiral galaxies selected from the SDSS-DR13/MaNGA database,
we examine the change of resolved properties of galaxies as the decrease of the global SFR.
We find that 
galaxies slow the speed of SF across their entire disks after they fall out of the SFMS,
with a decrease in galaxy center that is much quicker than that in outer disks.
Stellar mass also has a quick build-up at galaxy center, 
with a speed that is 0.4\,dex faster than that on disks farther out than $\sim$4\,kpc.
Galaxy center appears like a LINER host after the SF within the central 2\,kpc decreases to $\le 0.01\, \rm M_{\odot}\,yr^{-1}$.
From then on,
a hole in sSFR distribution becomes more significant, 
with the stellar component at galaxy center dominated by the old population,
which continuously expands outward as the global damping of SF.
The SFR profile transits from a central-peak function into a ``ring-like'' central-dip pattern, 
with a relatively active (but still faint) SF piling up at around $\sim$5\,kpc.

\begin{figure}
\plotone{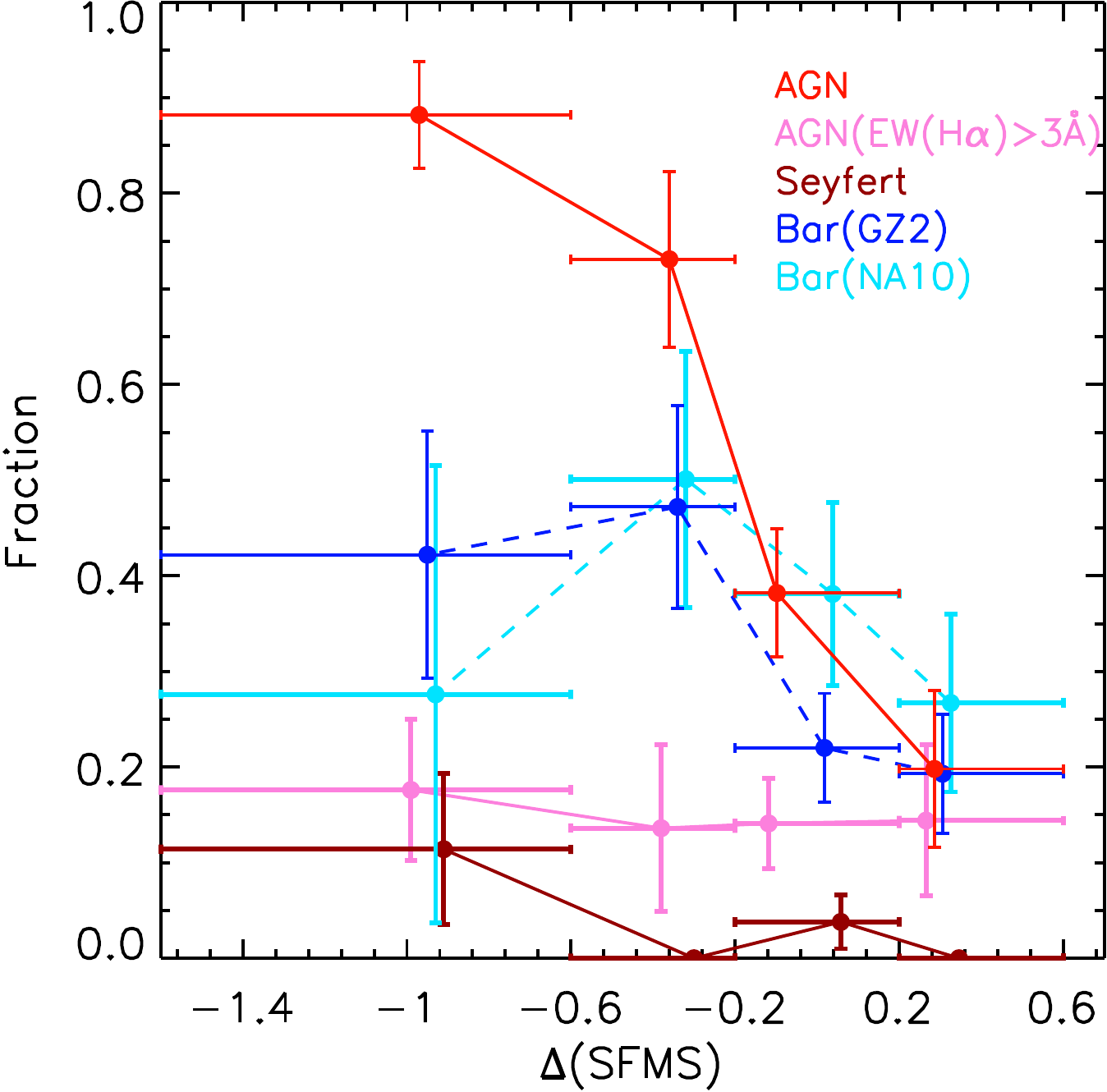}
\caption{Fraction of AGN-host and barred galaxies in a sample of different populations in terms of global star formation stage 
(from left to right: Red Sequence, Green Valley, Main Sequence, Starburst).
Galaxies with their central pixel belong to {\it LINER-like} or {\it Seyfert-like} regions 
in a resolved-BPT diagram of individuals, are classified as AGN hosts.
We adopt bar classification from two morphology catalogs, Galaxy Zoo 2 \citep{Willett13} and \citet{NA10}, 
125 and 72 matches are found respectively.
Errors shown in this figure are calculated from the bootstrapping method and the medians of mass are slightly moved for a clear distinction.
An increasing fraction of barred galaxies during the fading of star formation is suggestive from this plot (blue and cyan).
At the same time, an increase in AGN fraction based on resolved BPT diagram is also clearly shown (red).
However, this trend almost disappears when only Seyfert galaxies (dark red) or those with central $\rm EW(H\alpha)>3\AA$ (pink) are considered.
\label{fig6}}
\end{figure}

Our findings are consistent with the popular ``inside-out'' mode of quenching for massive galaxies,
in which central SF drops faster than outer disks.
Combined with the mass build-up in the central 2\,kpc before galaxies enter ``GV'' , 
our results agree with \citet{Fang13} in structure-quenching mechanisms \citep[see also][]{Tacchella15,Tacchella17}, 
which may be caused by stabilization of gas by the bulge/bar-driven process,
during which AGN feedback could also take a part.

To qualify this, we plot the evolution of the fraction of barred galaxies and AGN hosts
during the fading of global SF of disks in Figure~\ref{fig6}.
We adopt a morphology catalog from Galaxy Zoo 2 \citep{Willett13} and \citet{NA10} to classify barred galaxies in our sample.
Differences between the classification are mainly due to the deviation in sample coverage of these catalogs.
AGN hosts are galaxies with their central pixels classified as ``LINER-'' or ``Seyfert-like'' in resolved BPT diagrams (see Figure~\ref{fig4})
\footnote{Although composite objects could also be powered by AGNs, 
their numbers are small (3, 7, 1, and 0 in four groups, respectively) compared to either star-forming or AGN-like galaxies,
and could be counted into statistical errors in AGN fraction.}.
The increase of AGN fraction during the passive evolution is clearly seen from Figure~\ref{fig6}.
A significant increase in bar fraction after galaxies leave MS and fade into GV is also indicated
\footnote{The increase in the bar fraction from SFGs to quiescent galaxies is more prominent if we constrain the sample to galaxies with ${M_*}\leq 10^{11}\rm M_\odot$, 
which is from 0.11 for SB and 0.12 for MS, to 0.50 for GV and 0.43 for RS.}.

\subsection{Dynamical processes in inside-out quenching}
As our target galaxies are massive spirals that are not supposed to have experienced significant 
disruption in structure triggered by, e.g., major mergers,
secular evolution is an important mechanism for galaxy evolution and quenching.
Bars are suggested to play a central role in building the dense central component,
by driving gas inflow to trigger SF in galaxy center \citep{KK04, LC17}.
After the in-falling gas has all been consumed by stars,
a strong bar ``sweeps'' out star forming resources or increases the gas random motions within the corotation radius \citep{Kh17, Spinoso17}, 
leaving an ``star formation desert'' within the central $\sim$2\,kpc \citep{JP17}.

Compared with $\sim20\%$ and $\sim22\%$ bar galaxies found in {\bf SB} and {\bf MS},
half of our spirals in {\bf GV} and {\bf RS} are identified as barred galaxies in Galaxy Zoo 2 \citep{Willett13} and \citet{NA10}.
The increase of the bar percentage with the decrease of both global and inner-several-kiloparsec SF,
which is most prominent between MS and GV,
agrees with the ``bar-quenching'' mode
suggested in massive galaxies from both observation and simulation \citep[][see also \citealt{Kim17}]{Gavazzi15,Haywood16,Kruk18}.
Additionally, almost all of the last half central-red galaxies are inner lens/oval galaxies.
From a scenario in which bars dissolve into lenses\citep{Heller07}, 
the quenching of the central SF in these galaxies could also be driven by bars.
However, we did not find a significant enhancement of central SF caused by bar-triggered gas inflow before the suppression,
in barred galaxies of {\bf SB} and {\bf MS} compared to those unbarred,
which is observed in \citet{Kim17} and \citet{CT17}.
This may be caused by our limited barred sample (6 found in {\bf SB} and 12 found in {\bf MS}).
Bar strength could be another reason because there is barely any difference found between weakly barred and unbarred galaxies \citep{Kim17}.

The anticorrelation between the SF activity and bar fraction shown in Figure~\ref{fig6} 
is also consistent with theoretical studies in which 
bar formation is argued to be suppressed or delayed in gas-rich disks\citep[e.g.,][]{Ath13},
implying a possible relation between bar size/strength with on-going SF or optical color, 
as pointed out by \citet{Erwin18}. 
Similarly in observation, \citet{Cer17} has found an anticorrelation between H{\scriptsize\,I} gas richness and bar fraction,
agreeing with studies in which bars are more frequently found in massive and 
red early-type galaxies \citep[e.g,][]{Erwin05,NA10,Lee12}.
On the other hand, bars can also take a part in fueling gas into the galactic center and 
can trigger an enhancement of SF in situ \citep[e.g.][]{Ber07}. 
The lack of discovery of the high fraction of strong bar for SB/starbursts, which however was shown by \citet{JW12},
could result from the deviation in bar identification between studies.
Given that the bar strength in \citet{JW12} is parameterized in terms of ellipticity,
the contradiction could also be due to a potential dependence of bar ellipticity on central stellar mass, 
which is a natural result of the inverse correlation between 
bar ellipticity and central dynamical mass concentration found by \citet{Das03} based on CO observation.

The finding of a ring in SFR distribution between around 4-6\,kpc ($\sim2\,R_{\rm e}$) is noticeable in the last two stages of global SF.
This wide range shown in stacked 1D and 2D profiles resulted mostly from the scatter in galactic size.
While a bar triggering suppression of the central SF naturally lead to a ring-like SF in the outer part of a galaxy,
a combination of AGN feedback and bar-driven gas inflow also triggers an enhancement of SF and forming a dense ring at a finite radius \citep{Robichaud17}.
As \citet{Robichaud17} address, this scenario is rather a ``displacement'' than a ``suppression'' of SF,
because the gas is pushed to outer radii instead of thermalized or consumed.

A related high-redshift work \citep{Gen14} that also found a $\rm H\alpha$ ring argued for a different inside-out quenching mechanism for massive SFGs,
where a centrally peaked Toomre-{\it Q} distribution caused by the higher concentration of stellar mass
prevents further SF in galaxy center.
This ``gravitational quenching" or ``Morphological quenching" \citep{Martig09} has been highlighted 
in explaining the formation of red galaxies with quenched gas disks.
And SF is confirmed less efficient in bulge-dominated galaxies than in pure disks 
from both H{\scriptsize\,I} and $\rm H_2$ observation \citep[e.g.,][]{K89,Sain12}.
However, compared to the consistency in the effect of the bulge-related mechanism argued among the literature,
the role that bar leads in quenching is still concealed.
A further analysis with a more complete sample of the dynamics and the gas contents of barred galaxies is required to test the scenarios. 

The passive evolution of disks as shown in our work is related to but not exactly the 
``compaction'' suggested for quenching of high-redshift galaxies \citep[][and reference therein]{Dekel09,Zolotov15}.
Similar mechanisms require galaxies that are still on the main sequence experiencing a starburst to induce gas depletion and quenching afterward, 
which is also suggested in recent work by \citet{Ellison2018} based on their analysis of the gas-phase metallicity profile.
Though {\bf SB} is more metal-poor gas rich at galaxy center,
our finding of the higher $\Sigma(M_*)$ in the center of {\bf MS} than {\bf SB} galaxies shows that
starbursts may not be powerful enough to cause a compaction, 
either to deplete gas in galaxy center or to quench SF.
However, we should restate that 
compared to those extremes, which form most of stars in very short timescales,
our {\bf SB}s are more likely normal SFGs scattering above SFMS further than normal SFGs.
Additionally, accounting for the relatively short timescale of ``compaction'',
the $\Sigma(M_*)$ profile for {\bf SB} and {\bf MS} galaxies plausibly do not show significant difference.
Nevertheless, the faster mass growth in galaxy center compared to that in the outer disk for fixed $M_*$ certainly indicate a shrink in galaxy size (effective radius),
which could also be caused by dynamical reasons like bar-driven processes other than SF only.
A more detailed self-consistent analysis with a proper assumption in SFH is required for further studies.

\subsection{The puzzling role of AGNs in quenching SF}
Only 10 in 117 of our galaxies are classified as AGNs in the MPA-JHU catalog \citep{Tremonti04} based on the SDSS spectrum, 
including 1 {\bf SB}, 5 {\bf MS}s, 3 {\bf GV}s and 1 {\bf RS} galaxies.
However, from resolved analysis,
16 of 22 {\bf GV}s and 20 in 23 {\bf RS} galaxies show AGN-like emission in their central regions,
and 34 of them in total could be classified as LINERs.
Compared with the H{\scriptsize\,II} dominated ``Starburst'' and ``Main Sequence,'' 
this finding of the LINER-dominated other two groups qualitatively agree with the recently suggested ``LI(N)ER'' sequence of \citet{HL17}.
These galaxies are either classified as Type-2 AGN hosts \citep{CT17} or ``cLIER''s \citep{B17} in related studies.
However, because of their old underlying stellar population, 
and relatively extended spatial distribution, 
instead of low-luminosity activity nuclei,
the emission sources of the harder ionization field compared to the H{\scriptsize\,II} region are suspected to be 
low-mass hot evolved stars \citep[e.g.,][]{Bin94, Sarzi10, CF11, YB12, Papaderos13, Singh13}.

\begin{figure}
\plotone{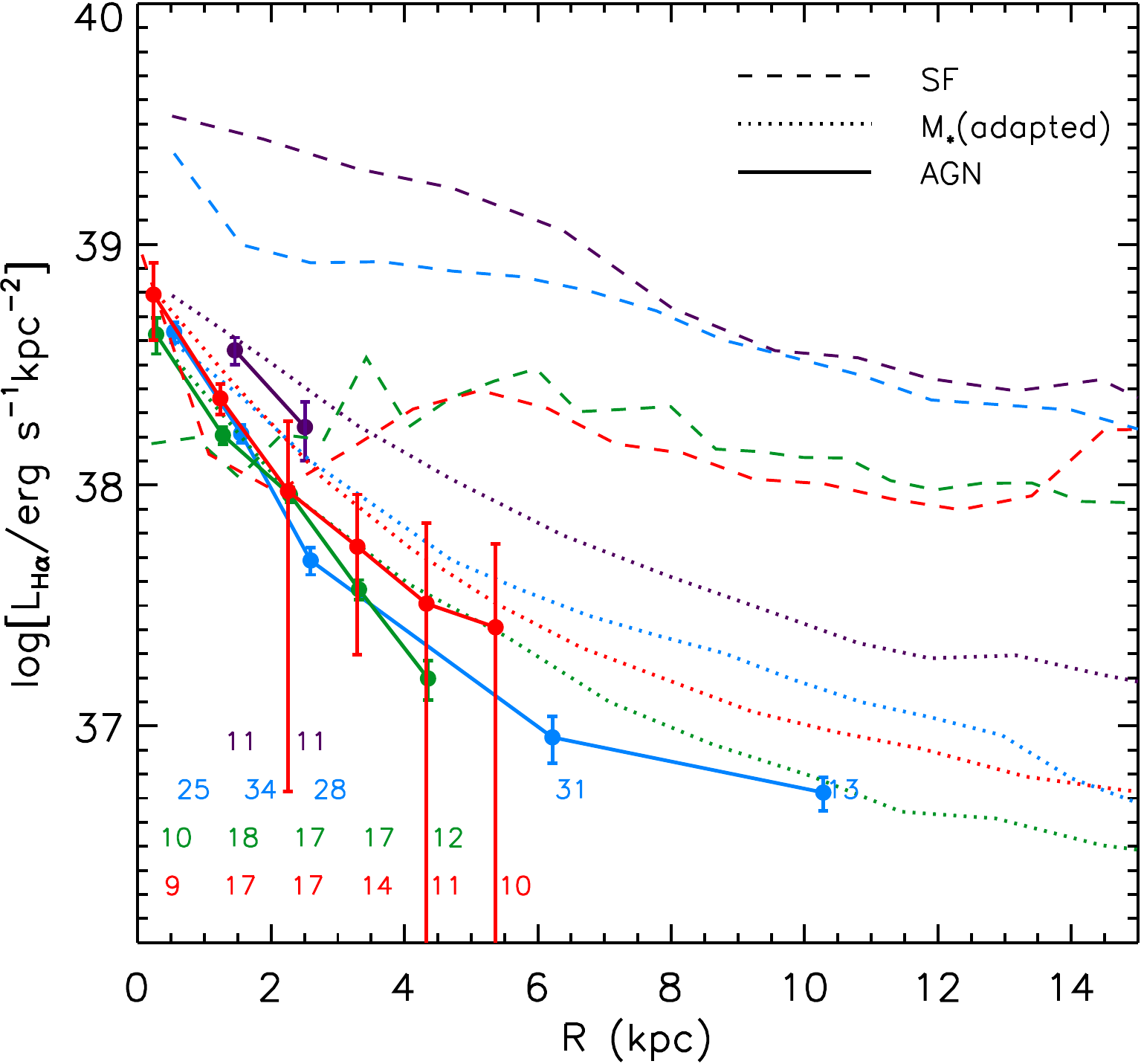}
\caption{Median $\rm H\alpha$ profiles of AGNs (solid lines) in comparison with that from star formation (dashed lines).
${\Sigma(M_*)}$ profiles are also shown as dotted lines with adaptive normalizations to be compared with the $\rm H\alpha$ profiles.
The color coding is the same as that in Figure~\ref{fig5}.
Only bins in which the number of galaxies stacked is not less than nine are shown.
Numbers in each bin are indicated above the $x$ axis.
H$\alpha$ emission from AGNs is based on the AGN-SF decomposition described in \S3.2.
Dust attenuation is corrected with the assumption of intrinsic ${\rm H\alpha/H\beta=3.06}$ \citep{Dong08}.
A central-peak pattern in AGN emission is clearly shown.
\label{fig7}}
\end{figure}

A mixture of low-luminosity AGN and old stellar component indeed could not be ruled out in red galaxies.
The universality of the ``LINER-like'' emission we find in the SF suppressed disk center 
agrees with the concentrated distribution of the old stellar population in red galaxies,
the emission from which is overshined by SF before a galaxy drops off the SFMS,
preventing itself from being distinguishable until young and massive stars fade away.
On the other hand, given the gently decreasing $\Sigma(M_*)$ profile from center to outside in each population,
the point-source-like central-peak pattern of the $\rm H\alpha$ profile contributed from pure ``AGNs''
(from the empirical AGN/SF decomposition) as shown in Figure~\ref{fig7}
implies a plausible existence of ``real'' LINER emission within the central 2\,kpc,
and could hardly be explained by an excess of the population of hot evolved stars 
at galaxy center \citep[also see the discussion in \S6.3 of][]{Ho08}.

A quick check of individual 2D EW($\rm H\alpha$) maps reveals that around one-fifth of our ``cLIERs'' should be true AGNs 
from their emission in central pixels
\footnote{The result does not change within 1$\sigma$ if we calculate the integrated $\rm EW(H\alpha)$ within a $R_{\rm e}$ or 2.3kpc 
(corresponding to 1'' in $z\sim0.13$) instead.} 
based on the criteria suggested by \citet{CF11}.
However, the increasing pattern of the AGN fraction disappears if we throw out the ``retired galaxies'' defined above or if only Seyferts are considered (Figure~\ref{fig6}).
Thus despite the plausible contribution of AGNs to the steep {$\rm H\alpha$} profiles,
the concentrated $\rm H\alpha$ emission in {\bf GV} and {\bf RS} could also partly be seen as a combined result of 
the depression in central H{\scriptsize\,II} emission (dashed line in Figure~\ref{fig5}) and the centrally concentrated distribution of stellar mass,
i.e., an outcome of the competition between young and old populations.
In this case, the concentrated old population at galaxy center, i.e., bulge, 
could play a more important role than AGNs in triggering the quenching of an SFG,
given that most stellar mass at galaxy center has already been formed before a galaxy become {\bf GV}.
This process is related to the ``morphological quenching''\citep{Martig09} or ``gravitational quenching''\citep{Gen14} discussed in the last subsection.
Nevertheless,
given the well-known correlation between the mass of the black hole and that of bulge,
a {\bf GV} or {\bf RS} galaxy that has already built up its compact central component should undoubtedly contain a massive black hole.
Moreover, 
because $85\%$ (57 in 66) of nearby LINER host galaxies were found to have X-ray cores in a recent study \citep{She17},
we cannot rule out the effect of AGN feedback on SF given our high fraction of LINER hosts in nearly quenched spirals.

Indeed, besides the processes caused by central stellar bulge or bar,
our findings also agree with AGN-driven quenching in state-of-the-art simulations.
Given the high fraction of ``LINER-like'' host galaxies in contrast to the rare ``Seyfert'' hosts (Figure~\ref{fig6}),
LLAGN could take a more important role in affecting global SF in galaxies.
The low-accretion-rate mode is found to dominate the duty cycle of an AGN \citep[e.g.,][]{Ho08},
suggesting an important cumulative effect of the corresponding ``hot-mode feedback'' \citep{Yuan18} on SF.
Both energy and momentum are injected into the surrounding inter-stellar medium,
thermalizing, and/or diluting the materials for further SF.

Wind emitted from AGNs is found to play a more important role in quenching galaxy disks than radiation 
\citep[Figure 8 in][see also \citealt{Wein17a,Wein17b}]{Yuan18}.
Our finding of the suppression of SF in the central 2\,kpc region in LINER hosts is likely because of the kinetic
wind launched by the hot accretion flow \citep{Yuan15} in the ``hot feedback mode.''
\footnote{The $\rm\Sigma(SFR)$ peak we found for green and red galaxies within the central 1\,kpc of a disk 
is also predictable in quenching models of \citet{Wein17a} for a galaxy with central-peak gas distribution.}
Although there could be other mechanisms depleting cold gas and preventing gas inflow
when gas supply is still plentiful at galaxy center,
our results suggest a possible quenching mechanism of low-luminosity AGN feedback, 
if the emission source of ``LINER'' could be confirmed as a real AGN.

\acknowledgements
We thank the anonymous referee for their valuable comments and suggestions that helped to improve the manuscript.
We thank the discussions with Prof. Jing Wang, 
Prof. Martin Bureau, Prof. Michele Cappellari, Prof. Chris Lintott, 
and Dr. Zhi-zheng Pan.
The first author is grateful for the constant encouragement of Prof. Xian-zhong Zheng.
Y.P. acknowledges support from the National Key Program for Science and Technology Research and Development under grant No. 2016YFA0400702, and the NSFC grant No. 11773001.
H.F. acknowledges support from the National Science Foundation under grant AST-1614326.
Parts of this research were conducted by the Australian Research Council Centre of Excellence for All Sky Astrophysics in 3 Dimensions (ASTRO 3D), 
through project number CE170100013.
This research is also supported jointly by China National Postdoctoral Science Foundation,
China Scholarship Council and The International Centre for Radio Astronomy Research. 

Funding for the Sloan Digital Sky Survey IV has been provided by the Alfred P. Sloan Foundation, the U.S. Department of Energy Office of Science, and the Participating Institutions. SDSS-IV acknowledges
support and resources from the Center for High-Performance Computing at
the University of Utah. The SDSS website is www.sdss.org.

SDSS-IV is managed by the Astrophysical Research Consortium for the 
Participating Institutions of the SDSS Collaboration including the 
Brazilian Participation Group, the Carnegie Institution for Science, 
Carnegie Mellon University, the Chilean Participation Group, the French Participation Group, Harvard-Smithsonian Center for Astrophysics, 
Instituto de Astrof\'isica de Canarias, The Johns Hopkins University, 
Kavli Institute for the Physics and Mathematics of the Universe (IPMU) / 
University of Tokyo, Lawrence Berkeley National Laboratory, 
Leibniz Institut f\"ur Astrophysik Potsdam (AIP),  
Max-Planck-Institut f\"ur Astronomie (MPIA Heidelberg), 
Max-Planck-Institut f\"ur Astrophysik (MPA Garching), 
Max-Planck-Institut f\"ur Extraterrestrische Physik (MPE), 
National Astronomical Observatories of China, New Mexico State University, 
New York University, University of Notre Dame, 
Observat\'ario Nacional / MCTI, The Ohio State University, 
Pennsylvania State University, Shanghai Astronomical Observatory, 
United Kingdom Participation Group,
Universidad Nacional Aut\'onoma de M\'exico, University of Arizona, 
University of Colorado Boulder, University of Oxford, University of Portsmouth, 
University of Utah, University of Virginia, University of Washington, University of Wisconsin, 
Vanderbilt University, and Yale University.

\end{document}